# The superite phase and phase transition inducing multiscale solidification microstructures and segregations in steels

Xiaoping Ma[*1], Dianzhong Li[1], Zhuo Zhao*[2], Saichao Cao[3], Donghao Pei[1], Pei Wang[1], Yuxi Tao[1], Paixian Fu[1], Hongwei Liu[1], Xiuhong Kang[1]

*1 Shenyang National Laboratory for Materials Science, Institute of Metal Research, Chinese Academy of Sciences, Wenhua Road No. 72, Shenyang, 110016, P. R. China*

*2 Department of Materials Science and Engineering, University of Science and Technology Liaoning, Qianshanzhong Road No. 189, Anshan, 114051, P. R. China;*

*3 Shanghai Synchrotron Radiation Facility (SSRF), Shanghai Advanced Research Institute, Chinese Academy of Sciences, Zhangheng Road No. 239, Shanghai, 201210, P. R. China*

**Based on classical concept, solidification of alloys is a direct transition from liquid phase to solid phase, by which dendrites and dendritic segregation are produced. Through in-situ and real time morphology observation and XRD test during solidification of three steels, a new superite phase featured as statistically oriented tiny structures was identified, and a general liquid-superite-solid phase transformation process is revealed. In the early solidification stage, the liquid alloys transit to dendrites composed of superite phase. Initiated from the boundaries of dendritic arms or dendrite grains, the superite phase transits to austenite grains within an initial dendritic arm, and expels solute elements to the residual superite**

* Xiaoping Ma, Phone: 86-024-83970106, Contact e-mail: xpma@imr.ac.cn
* Zhuo Zhao, Contact e-mail: zhaozhuo@ustl.edu.cn

**phase. Mixed multi-phase microstructures are subsequently produced from the residual enriched superite phase. Here, although three steels exhibit different phase proportion and phase constitution in the superite-solid transition, they all follow above general transition mode. Multiscale microstructures and segregations are produced in the transition from superite to solid. These new findings change the basic understanding about the solidification of alloys, rediscover the formation mechanism on segregations and multiscale solidification microstructures, including dendrite pattern, solid dendritic arm, dendritic segregation, the mixed multi-phase microstructures, eutectic, inclusions and precipitate. These new findings are also crucial to the control of solidification microstructures and segregation in metals.**

Dendritic solidification is an important issue in material science. Dendrites morphology and dendritic segregation significantly influence the mechanical and processing properties of alloys [1, 2]. The dendritic arm spacing can influence the strength and ductility [3, 4], and the dendritic segregation can influence the amount and distribution of precipitates and detrimental phases, which relate to the reliability and performance of alloys [5, 6]. The formation mechanism for dendrites morphology and dendritic segregation is a key scientific topic in solidification field. According to classical solidification principle, as the results of anisotropic crystal growth, interface stability and the gradient of temperature and compositions, the solid grows to form a dendritic morphology [7-9], and the dendritic segregation is produced by the solute partition between solid and liquid in the solid growth process [10, 11]. The basis lies in that solidification of alloys is a direct solid growth process from liquid, by which dendrites and dendritic segregation are produced. Although it was realized that a semi state exists between liquid and solid in atomic crystal nucleation [12-14], the semi state was previously not considered in the growth process of dendrites. Our previous investigations about Al-Cu alloy and steels indicated that the traditional dendritic arm should be composed by a semi-solid matrix rather than pure solid [15-19]. Especially, it was in-situ and real time observed that multiscale segregation patterns sequentially emerged on the surface of dendritic arms [20, 21]. Thus, a new solidification principle based on semi-solid matrix is proposed in the term of discrete solidification [22]. In a semi-solid matrix, the temperature gradient can induce gradients of solid fraction, solid compositions and liquid compositions. Above gradients will drive the mass transfer to produce multiscale segregation patterns and solidification microstructures. Although above principle for discrete solidification was elaborated by modelling and some insufficient in-situ observation [22], only the tip of the iceberg was revealed by limited investigations about the evolution of the semi-solid matrix. Many puzzles about semi-solid matrix remain to be clarified with strong evidences. Does the semi-solid matrix have a new and independent matter structure or just a mixture of liquid phase and known solid phase? What is the whole evolution process in liquid-semi solid matrix-solid transition, which producing multiscale solidification microstructures and segregations? Is there a general mode in the transition from semi-solid matrix to solid in different alloys? Whether solidification microstructures and segregations can be controlled by the transition from semi-solid matrix to solid? Through the in-situ and real time observation on the solidification morphology evolution on sample surface by laser confocal scanning microscope, in-situ and real time XRD test by synchrotron radiation facility and the solidification microstructures observation both on the sample surface and in the sectioned sample, above questions are clarified in this paper by the investigation on three steels, including M50 bearing steel, 310s stainless steel and 321 stainless steel.

1. The solidification of M50 steel

In the in-situ morphology evolution observation and XRD test during solidification of M50 bearing steel, the sample was melted and then solidified in a cooling rate of 1℃/s. the key points of morphology evolution and XRD results are shown in Figure 1. The detailed morphology evolution process is provided in the supplementary video S1. The initial dendrites emerged and grew from liquid below 1415℃. A few amplified initial dendritic arms at 1350℃ are shown in Figure 1 (a). The XRD pattern and peaks of pure liquid are shown in Figure 1 (i) and (m). The XRD pattern and peaks of liquid and initial dendrites in the Figure 1(a) are shown in Figure 1 (j) and (n). Negligible δ ferrite appeared with very weak peak intensity, which cannot be actually observed in Figure1 (a). Actually, the initial dendritic arms in Figure1 (a) are composed by a new and independent phase termed as superite, which is featured as a large amount of statistically orientated tiny structures with its unique XRD pattern shown in Figure 1 (j). The key difference between the liquid and superite lies in the statistically orientation. The orientation of a large amount of tiny structures is random in liquid with uniform intensity distribution in the hallo ring of liquid, while the orientation of a large amount of tiny structures is statistically orientated in superite with unique intensity distribution as indicated in Figure 1 (j). It should be highlighted that after the formation of superite, the intensity of diffraction remain a low level comparable with the liquid. Subsequently, as shown in Figure 1 (b)-(d), these superitic dendritic arms abruptly began the transition process to austenite at 1349.9℃, which induced some solidification morphologies within dendritic arms and a temperature jump by the latent heat of phase transition. The XRD pattern and peaks corresponding to the Figure 1(c) shown in Figure 1 (k) and (o) verifies the superite to austenite transition, in which the austenite with many orientations appeared with increasing intensity and the XRD pattern of superite became weak. As this sample was further cooled, accompanied with the transition from superite to austenite, some austenite grain boundaries, spot segregations and tiny dendritic segregation zone successively emerged on the surface of traditional dendritic arms, as shown in Figure 1 (e)-(g). The 3D morphology of the tiny dendritic segregation zone is shown in Figure (h). Some concavities exist within the tiny dendrite and especially on the envelope of the tiny dendrite. The XRD pattern and peaks after fully solidification are shown as Figure 1 (l) and (p), in which the superite disappeared and the intensity of austenite strongly increased.

After solidification, the grain boundaries, spot segregations and tiny dendritic segregation zone were further analyzed by the laser confocal scanning microscope, EBSD and EDS mapping. The tiny dendritic segregation zone is composed of austenite with different orientations, while the rest part of traditional dendritic arms is composed of martensite, as evidenced in Figure 2 (a)-(d). The full compositions (wt.%) of the tiny dendrite zone are Fe 82.94, C 7.94, Si 0.38, Mn 0.12, Cr 3.66, Mo 3.24, V 1.33, Ce 0.29, S 0.10, and the full compositions (wt.%) of the black spot are Fe 67.74, C 9.44, Si 0.68, Mn 0.95, Cr 3.02, Mo 2.65, V 1.11, Al 1.31, Zr 1.21, La 2.62, Ce 3.93, O 4.58, S 0.76. Although the main element in the tiny dendrite and in the spot segregation is Fe, the boundary and tiny dendrite are comparatively enriched with C, Mo, V, Ce and S elements, and the spot are comparatively enriched with Al and O elements, as shown by the EDS mapping in Figure 2 (e)-(l). The segregated spots, grain boundary and tiny dendritic segregation zones with inner dendrites in-situ observed on the surface of traditional dendritic arms reflect the solute expelling during the transformation from superite to austenite. The interior solidification microstructures in the sectioned sample are shown in Figure 2(m)-(t). The traditional dendritic arm is composed of some small grains, and the interdendritic position is composed of the eutectic. As shown in Figure 2 (o)-(t), the small grain is composed of the martensite, which should transit from austenite in the quenching process. The C, Mo and V elements segregate on the grain boundaries. The eutectic is a mixture of $Mo_2C$, $V_2C$, austenite and martensite. In the eutectic, the elements of C, V, Mo and Cr are enriched. The grain boundaries shown in in-situ observation should be equivalent to the grain boundaries within traditional dendritic arms in Figure 2 (m) and (n). In the interior of dendritic arms, because the solidification of superite expelled the solutes into the grain boundaries and final interdendritic positions, the austenite grains were produced within dendritic arms.

The transition from superite to austenite grains within initial dendritic arms was further supported by a supplementary sample of M50 steel solidified in a cooling rate of 1°C/s from 1500°C to 1370°C and then quenched to the room temperature by the liquid GaInSn alloy. The most of superite within initial dendritic arms transited to many austenite grains, and some residual superite transited to δ ferrite, as shown in Figure S2.

Above results describe the following solidification evolution. The liquid transited to superite in the dendritic pattern. As the superite phase was undercooled in the cooling process, the superite within traditional dendritic arms undergone abrupt phase transition to austenite with obvious latent heat release. The fraction increase and distribution of austenite produced the solidification microstructures within dendritic arms. As the further growth of austenite grains, the segregated elements (Cr, Mo and V typically) were ejected from the austenite grains to the grain boundaries and interdendritic positions. Finally, the eutectic was produced in the interdendritic positions. On the sample surface of dendritic arms, as the growth of austenite grains, partial austenite grain boundaries emerged on the surface with the enrichment of expelled C, Mo, V, Ce and S elements. The expelled C, Mo, V, Ce and S elements also segregated in some spots and tiny dendritic zones. After full solidification, the differentiated zones with and without solute enrichment all transited to austenite. In the quenching process, the austenite in enriched tiny dendritic zones was reserved, while the rest austenite without solute enrichment transited to martensite. All the solidification microstructures and segregations of M50 steel are dependent on the transition from superite phase to solid phase.

2. The solidification of 310s steel

In the in-situ morphology evolution observation and XRD test during solidification of 310s steel, the sample was melted and then solidified in a cooling rate of 0.1°C/s. The key points of morphology evolution and XRD results are shown in Figure 3. During the in-situ observation on the 1# sample, as the temperature decreased to 1369°C, the liquid transited to superite phase in the pattern of dendritic arms, as shown in Figure 3(a). The XRD pattern and peaks of pure liquid are shown in Figure 3 (m) and (q). The XRD pattern and peaks of liquid and initial dendrites in the Figure 3(a) are shown in Figure 3 (n) and (r). Negligible δ ferrite appeared with very weak peak intensity, which cannot be observed in Figure3 (a). Actually, the dendritic arms in Figure 3(a) are composed by a new and independent phase termed as superite, which is featured as a large amount of statistically orientated tiny structures with its unique XRD pattern shown in Figure 3(n). Subsequently, this superite phase within initial dendritic arms abruptly transited to some austenite grains, and some small dimples were induced within dendritic arms by shrinkage, as shown in Figure 3(b). Induced by the latent heat released along with the phase transition, a temperature jump appears, as shown in Figure 3 (d). The XRD pattern and peaks corresponding to the Figure 3(b) shown in Figure 3 (o) and (s) verifies the superite to austenite transition, in which the austenite with many orientations appeared with increasing intensity and the XRD pattern of superite became weak. As the further decrease of temperature, the austenite grains were obviously coarsened. When the temperature decreased to 1330°C, a lot of $(Si, Al)_xO_y$ particles appeared on the surface of dendritic arms. The coarsened austenite and $(Si, Al)_xO_y$ particles are shown in Figure 3 (c). This detailed evolution process is shown in the supplementary video S2. The XRD pattern and peaks after fully solidification are shown as Figure 3 (p) and (t), in which the superite disappeared and the intensity of austenite strongly increased. During the in-situ observation on the 2# sample, as the temperature decreased from 1420°C to 1404°C, the liquid transited to superite phase in the pattern of dendritic arms, as shown in Figure 3(c). Subsequently, the superite phase within traditional dendritic arms abruptly transited to some solidifying grains. Induced by the latent heat released along with the phase transition, a temperature jump appeared, as shown in Figure 3 (f)-(h). The time points of Figure 3 (e)-(g) are labelled in Figure 3 (h). This detailed evolution process is shown in the supplementary video S3. At about 40 seconds after the abrupt transition to small austenite grains, this sample was quenched in the cooling rate of 100°C/s from 1400°C to the room temperature. During the in-situ observation on the 3# sample, as the temperature decreased from 1420℃ to 1390℃, the liquid transited to the superite phase in the pattern of dendritic arms. The superite phase was quenched in the cooling rate of 100℃/s from 1390℃ to the room temperature. During quenching, a quick phase transformation process happened at 1325.9℃, as shown in Figure 3 (i)-(l). Some superite phase transited to austenite grains, while some other superite phase transited to a mixture of rod-like or spot-like ferrites and austenite base, as shown in Figure 3(k). Because the cooling is too fast in the quenching process, the temperature jump didn't appear along with the abrupt solidification of semi-solid phase, as shown in Figure 3 (l). The detailed evolution process is shown in the supplementary video S4.

After solidification, the microstructures on the surface of three samples were further analyzed by the laser confocal scanning microscope, SEM, EBSD and EDS mapping. The morphology and compositions of the $(Si, Al)_xO_y$ particles in 1# sample are amplified in Figure 4(a)-(d). The small austenite grains with different crystal orientation transited from superite within traditional dendritic arms in 2# sample are reserved by quenching process, as evidenced by EBSD mapping in Figure 4 (e)-(h). Figure 3(i) shows the step morphology or punctate morphology on the surface of austenite grains with different orientations, which indicates the step growth of austenite during superite to austenite phase transition. The austenite grains and mixture of rod-like or spot-like ferrites and austenite base transited from superite phase in 3# sample are shown in Figure 4 (j)-(m). As shown in Figure 4(k), the ferrite is enriched with Cr but short of Ni, while the austenite is enriched with Ni but short of Cr. As evidenced in Figure 3(k)-(m), the austenite base in the mixture may have different orientation, and the step growth during phase transition induced step morphology or punctate morphology on the surface of austenite grains. The interior solidification microstructures in the 1-3# samples are shown in Figure 4 (n)-(p), (r)-(z), and (α)-(γ), respectively. The solidification microstructures in the 1# samples are composed by some coarse austenite grains and some zones mixed by austenite base and granular δ ferrite, as shown in Figure 4 (n)-(p). The solidification microstructures in the 2# and 3# samples are composed by some dark strip (with porosities), some white zones (austenite grains) and some dark zones (mixed austenite and δ ferrite). Indicated by the morphology of dark strips and by porosities in dark strips, these dark strips should be interdendritic positions of original dendritic arms. Both the dark strip and white zone are composed of austenite grains with different orientations, as evidenced by EBSD mapping in Figure 4(s)-(u). The microstructures in the dark zones are composed by some rod-like δ ferrite mixed in the austenite grains, as shown in Figure 4(v)-(z). In which, the δ ferrite is enriched with Cr but short of Ni, while the austenite is enriched with Ni but short of Cr, as shown in Figure 4(y) and (z). Notedly, the size, amount and distribution of austenite grains and mixed microstructures differ apparently in three samples. Generally, the distribution of grains and mixed microstructures inherit initial dendritic pattern in the 2# and 3# samples. The amount of mixed microstructures is much more in 2# and3# samples than in 1# sample. The compositions of different zones in three samples are shown in Figure 4(q). Generally, the dark strips are Cr depleted and Ni enriched, while the dark zones are Cr enriched and Ni depleted. During solidification of steels, it has been proved by thermodynamic calculation and experimental test that Ni is enriched in the austenite, but Cr is enriched in the liquid [23]. Thus, in the chronological sequence, the dark strips, the white zones and the dark zones transited sequentially from superite phase. The in-situ observation also approved that the white zones (austenite grains) transited earlier, and the dark zones (mixture of austenite and δ ferrite) transited later at a lower temperature. Above interior microstructures coincide well with the in-situ observation on the sample surface. The formation and coarsening of austenite grains in the white zones was in-situ observed in the 1# sample. The formation of austenite grains in the white zones was in-situ observed in the 2# sample. The formation of austenite grains in the white zones and the mixture of austenite and δ ferrite in the dark zones was in-situ observed in the 3# sample.

Above results describe the following solidification evolution. The liquid alloy first transits to a superite phase in a dendritic pattern. Initiated from the original interdendritic positions, the superite phase sequentially transits to austenite grains by step growth style, which produces the dark strips and the white zones. Simultaneously, the Cr element is expelled out of austenite, while the Ni element is enriched in the austenite. The residual superite phase transits to a mixture of austenite and δ ferrite in the later, which constitutes the dark zones. Within the dark zones, the Cr element is expelled out of austenite, while the Ni element is enriched in the austenite. Because the transition from superite phase to the austenite grains happened quickly in the quenching process of the 3# sample, the boundary between white zones and dark zones is interlaced. Some austenite grains can't converge completely, and some residual superite phase between separate austenite grains transit to a mixture of austenite and δ ferrite, as shown in Figure 4 (β). To the contrary, in the slow cooling after the transition from the superite phase to austenite in the 1# sample, the austenite grains obviously coarsened, and only a little of residual superite phases transit to sparse dark zones. Meanwhile, most trace elements, such as Al and O, were expelled into the superite phase on the surface of the 1# sample, which induced the (Si, $Al)_xO_y$ compound particles. Most important conclusions are that the final solidification microstructures and segregations in 310s steel are directly produced by the superite phase transformation. The abrupt progress of superite to austenite phase transformation happened in a short period of time but played a key role in the formation of solidification microstructures and segregations. By controlling the superite phase transformation, the solidification microstructures and segregations can be significantly influenced, as evidenced by the comparation in three samples.

The validity of the semi-solid phase can also be proved by the thermodynamic analysis. If the original dendritic arm is composed of the pure austenite, the Cr element will not be expelled out of austenite after its formation, because the equilibrium Cr content increases as the decrease of the temperature, as shown in Figure S1(e). However, experimental results approve the expelling of Cr element out of austenite. Besides, if the original dendritic arm is composed of the austenite, the dendritic arm formed in the early solidification stage should not contain significant Al and O elements to induce plenty of $(Si, Al)_xO_y$ particles on the surface of initial dendritic arms. Only if the original dendritic arm is composed of superite phase, the trace elements, such as Al and O, can be expelled from the superite phase to the surface of original dendritic arms. Therefore, the initial dendritic arms must be composed of the superite phase.

3. The solidification of 321 steel

In the in-situ morphology evolution observation and XRD test during solidification of 321 steel, the sample was melted and then solidified in a cooling rate of 2°C/s. The key points of morphology evolution and XRD results are shown in Figure5. From 1423°C, the superite dendrites emerged and grew from the liquid. From 1365°C, the transformation from superite to austenite was in-situ observed. As shown in Figure 5 (a)-(c), some granular or rod-like austenite blocks successively emerged within initial superite dendritic arms. Along with the successive emergence of austenite blocks, a temperature variation appears, as shown in Figure 5 (d). The detailed evolution process is shown in the supplementary video S5. The XRD pattern and peaks of pure liquid are shown in Figure 5 (e) and (i). The XRD pattern and peaks of liquid and initial dendrites in the Figure 5(a) are shown in Figure 5 (f) and (j). Negligible δ ferrite appeared with very weak peak intensity, which cannot be observed in Figure 5(a). Actually, the dendritic arms in Figure 5(a) are composed by the superite phase, which is featured as a large amount of statistically orientated tiny structures with its unique XRD pattern shown in Figure 5(f). Subsequently, this superite phase within initial dendritic arms quickly transited to some austenite blocks, as shown in Figure 5(b) and (c). Induced by the latent heat released along with the phase transition, a temperature fluctuation appears, as shown in Figure 5 (d). The XRD pattern and peaks corresponding to the Figure 5(b) shown in Figure 5 (g) and (k) verifies the superite to austenite transition, in which the austenite with many orientations appeared with increasing intensity and the XRD pattern of superite became weak. After fully solidification, the XRD pattern and peaks are shown as Figure 5(h) and (l), in which the superite disappeared and the intensity of austenite strongly increased.

The microstructures on the sample surface reserved after solidification are shown in Figure 5 (m). As shown in Figure 5 (n)-(t), the emerged austenite blocks are Ni enriched with different orientation, and the residual superite transited to δ ferrite enriched with C, Cr, Ti and Nb elements. Interestingly, the Ti, Nb and C elements further distribute in the residual superite phase, resulting in alternate small austenite and δ ferrite phases among large austenite blocks. It is also noted that some austenite blocks stretch over neighboring initial dendritic arms. As shown in Figure 5 (u)-(β), the interior solidification microstructures of the 321 stainless steel are composed by a network of austenite and some enclosed zones mixed by dendritic austenite blocks and interstitial δ ferrites. Indicated by porosities in the austenite network, as highlighted by red circles in Figure 5 (u), the austenite network should firstly transit from boundaries of original dendrite grains. Some austenite blocks and interstitial ferrites in a labelled square are amplified in Figure 5 (x). As shown in Figure 5 (y)-(β), the austenite is enriched with Ni element but short of Cr element, while the δ ferrite is enriched in Cr element but short of Ni element. The microstructures in the zones mixed by austenite blocks and interstitial δ ferrite are equivalent to the microstructures in in-situ observation.

Above results indicate the following solidification evolution. The liquid transits to superite phase in the dendritic pattern. On the sample surface, some superite phase within initial dendritic arms transited to the granular or rod-like austenite with different orientations, and the segregated elements (C, Cr, Ti, Nb) were ejected to the residual superite phase. As some residual superite phase further transited to small austenite with different orientations, the segregated elements (C, Ti, Nb) were further ejected to the residual superite phase, which finally transited to the δ ferrite. In the inner sample, the superite phase on the boundary of dendrite grains transited to a network of austenite, and the rest superite phase transited to a mixture of austenite and δ ferrite. The expelling of Cr element from austenite during the phase transition strongly supports the validity of the superite phase. If the original dendritic arm is composed of pure austenite, the Cr element will not be expelled out of new austenite blocks, because the equilibrium Cr content increases as the decrease of the temperature, as shown in Figure S1(f). The phase transformation from superite phase to austenite/δ ferrite completely eliminated the original dendritic arms in the final solidification microstructures. The phase transformation related with superite is imperative to understand and control solidification microstructures and segregations.

Some supplementary experimental evidence can further support the validity of the superite phase. This supplementary sample was solidified in a cooling rate of 2℃/s from 1500℃ to 1400℃, and was subsequently quenched to the room temperature in a cooling rate of 100℃/s. The austenite and δ ferrite formed within the original dendritic arm in the quenching process, which detailed solidification process is shown in the supplementary video S6. If the original dendritic arm is composed of δ ferrite, significant segregation should exist between the center and boundary of original dendritic arm, the transition from δ ferrite to austenite during quenching should not produce obvious segregation pattern of Ti element within austenite, the transition from δ ferrite to austenite during quenching should also not produce obvious partition of Cr, Ni and Mo elements between austenite and residual δ ferrite, and the residual δ ferrites should show identical crystal orientation. However, above deductions based on the assumption that the original dendritic arm is composed of δ ferrite are robustly disproved by the experiment results, as shown in Figure S3 (e)-(h), (i)-(m), (n)-(s). Therefore, the original dendritic arm must be composed of the new superite phase.

4. The general superite transition mode inducing multiscale solidification microstructures and segregations

Through above investigation and analysis on three steels, the understanding about solidification of steels is changed from a liquid-solid transition to a liquid-superite-solid transition. A new superite phase and related phase transformation processes are revealed. Despite some differences in details, this phase transformation process during solidification of three steels follows a unified mode, as sketched in Figure 6 (a). During solidification, the liquid firstly transits to dendrites composed of superite phase with initial microsegregation. Subsequently, either initiated from the boundary or center of dendritic arms, the superite phase sequentially transits to austenite grains. Along with the superite to austenite transition, the latent heat is released, and the solute elements are expelled from austenite into the residual superite phase, which contributes to the evolution of microsegregation. The residual superite phase finally transits to multi-phase microstructures. Figure 6(b) sketches the transformation process from superite phase to solid phase in M50 steel. In the first step, most superite phase transits to austenite grains, and the solute elements are expelled from the austenite grains into the grain boundaries and the residual superite phase. In the second step, the residual superite phase transits to multi-phase microstructures. The multi-phase microstructures produced on the surface of the sample are two kinds of austenite with different compositions, and the multi-phase microstructures produced in the interior of the sample are austenite and carbide. Most superite phase transited to austenite grains, and only a small amount of superite phase transited to the eutectic. Figure 6(c) sketches the phase transformation process from superite phase to solid phase in 310s steel. In the first step, the superite phase on the boundary of original dendritic arms transits to austenite grains, and the solute elements are expelled from the austenite into the superite phase. In the second step, some neighboring superite phase transits to austenite grains, and the solute elements are further expelled from the austenite into the residual superite phase. The compositions of austenite formed in the first and second steps are different. The transited austenite grains may become coarse depending on cooling rate. In the third step, the residual superite phase transits to solid grains composed of austenite and δ ferrite. Both the proportions of superite phase transited to the austenite grins and the multi-phase microstructures are significant. Figure 6(d) sketches the phase transformation process from superite phase to solid phase in the 321 stainless steel. A small amount of superite phase on the boundaries of dendrites transited to a network of austenite, and the majority of superite phase transited to multi-phase microstructures composed of austenite and δ ferrite. In summary, depending on the amount of solute expelled in the superite to austenite transition, three steels show different proportion and phase constitution in the multi-phase microstructures transition. However, three steels follow a general phase transformation mode as sketched in Figure 6(a).

The phase transformation from superite to austenite and multi-phase microstructures is a crucial process, by which multiscale solidification microstructures and segregation are produced. Some superite with solute depletion transited to austenite grains, and some solute enriched superite transited to multi-phase microstructures. This phase transformation can be controlled by the cooling process, and influence the amount and distribution of the austenite grains and the multi-phase microstructures, as evidenced by the results of 310s steel. Reformed or covered by the phase transformation from superite to solid, some initial dendritic patterns may completely annihilate in the final solidification microstructures. However, because the phase transformation from superite to solid tends to initiate from the boundaries of original dendritic arms or dendrite grains, the original dendritic pattern can also inheritably influence the final solidification microstructures and segregation distribution, such as the austenite network in the 321 steel. The phase transformation from superite to solid is a crucial process for the formation of microsegregation. The microsegregation is initially induced by the solute partition between the superite and liquid in the formation of dendrites. As the further solute partition during the transition from superite to austenite grains, the solute elements are expelled from austenite and enrich in the residual superite phase, which contributes to the further evolution of microsegregation. It should be noted that, because the transition from the superite to solid phase occurs at a much lower temperature, contrary to traditional understanding, the microsegregation undergoes significant evolution at a much lower temperature. Developing a clear picture of phase transformation from superite to solid can clarify the formation mechanism for solidification microstructures and microsegregation, and can help determine the optimum conditions necessary for the effective control on solidification microstructures and microsegregation by a previously omitted phase transition process.

**Acknowledgements** This work is supported by the General Program of National Natural Science Foundation of China (Grant No. 52271041) and SSRF proposal No. 2024-SSRF-PT-508818. The authors acknowledge the beamline BL12SW (http://cstr.cn/31124.02.SSRF.BL12SW) of Shanghai Synchrotron Radiation Facility (SSRF) in China for the experimental support. The Ultra-High Temperature Infrared Heating Furnace (IH1600-SR) used in in-situ HE-XRD test by synchrotron radiation facility is supported by the GoGo Instruments Technology (Shanghai) Co., Ltd. The authors thank Professor Xinzhong Li in Soochow University and Professor Xue Liang in Shanghai University for inspiration in investigation.

**Author Contributions** X.P. Ma propose the concept of superite phase and phase transition mechanism, performed all in-situ experiments, solidified sample characterization, result analysis, mechanism establishment and wrote the manuscript. D.Z. Li supported above investigation on superite phase, and contributed to the research facilities and financial support. Z. Zhao performed the laser confocal microscope in-situ and real time morphology observation during solidification. S.Ch. Cao and D.H. Pei performed the in-situ and real time HE-XRD test by synchrotron radiation facility during solidification. P. Wang prepared the raw material of 310s stainless steel. Y.X. Tao, P.X. Fu, H.W. Liu and X.H. Kang contributed to the discussion and revision on manuscript.

**Competing interests** The authors declare no competing financial interests.

## Methods

A laser confocal scanning microscope (VL2000DX-SVF18SP, Yonekura Manufacturing Corporation, Japan) with a purple laser diode (wavelength of 405 nm) was employed for in situ morphology observation. The samples were machined into a disc with the dimension of Φ 7.5×2.5 mm. Then, the specimens were mechanically ground, mirror polished, positioned in alumina crucibles, and immediately subjected to the specific thermal cycle under high-purity argon in the furnace. During in-situ observation, we firstly use a large scope to observe the emergence of initial dendrites. Once the traditional dendrites appear, we change to a small scope to observe several fixed dendritic arms. Successive images about the microstructural evolution during solidification were in-situ and real time recorded and collected by a charge-couple device (CCD) camera of the laser confocal microscope, and the cooling curves were recorded by a thermocouple. The sample of M50 bearing steel (with the composition wt.% of C 0.77, Si 0.20, Mn 0.17, Cr 4.02, Mo 4.30, V 0.97 and Fe balanced) was melted and held at 1500 ℃ for 5 minutes. This sample weas solidified in the cooling rate of 1℃/s from 1500℃ to 1000℃, and was subsequently quenched to the room temperature in a cooling rate of 10℃/s. Three samples of 310s stainless steel (with the composition wt.% of C 0.052, Si 0.96, Mn 1.48, Cr 25.0, Ni 20.0 and Fe balanced) was melted and held at 1500 ℃ for 5 minutes. The 1#, 2# and 3# samples were respectively solidified in the cooling rate of 0.1℃/s to 1300℃, 1400℃ and 1390℃, and were subsequently quenched in the cooling rate of 100℃/s to the room temperature. The samples of 321 stainless steel (with the composition wt.% of C 0.023, Si 0.53, Mn 1.60, Cr 18.7, Ni 9.63, Ti 0.25, Mo 0.16, Cu 0.10 and Fe balanced) were melted and held at 1500 ℃ for 5 minutes. The sample was solidified in a cooling rate of 2℃/s from 1500℃ to 1300℃, and was subsequently quenched to the room temperature in a cooling rate of 100℃/s. A supplementary sample weas solidified in a cooling rate of 2℃/s from 1500℃ to 1400℃, and was subsequently quenched to the room temperature in a cooling rate of 100℃/s. After solidification, the solidification microstructures on the surface and in the interior of the samples were observed by the optical microscope and SEM. The segregation distribution within microstructures was tested by the EDS mapping. The phase distribution and crystal orientation in the solidification microstructures were tested by EBSD.

A supplementary sample (Φ 4×10mm) of M50 steel was melted in an alumina crucible by an induction furnace with the Ar protection, and was solidified in a cooling rate of 1℃/s from 1500℃ to 1370℃ and then quenched to the room temperature by the liquid GaInSn melt. After solidification, the solidification microstructures in the interior of the samples were observed by the optical microscope and SEM. The segregation distribution within microstructures was tested by the EDS mapping. The phase distribution and crystal orientation in the solidification microstructures were tested by EBSD.

The in-situ HE-XRD experiments were conducted at beamline BL12SW of the Shanghai Synchrotron Radiation Facility (SSRF) in China. A monochromatic X-ray beam of energy 98.2974keV was used, with a beam size of 500×500μm$^2$. The diffraction signals were recorded in a frequency of 2s per frame with a two-dimensional detector (X-Panel 4343a FQI-H, 4288×4288 pixels, each pixel 100×100μm2), using an exposure time of 1s per frame. A $CeO_2$ standard sample was used to calibrate the sample-to-detector distance and detector orientation prior to testing. During test, the samples of M50 steel and 321 steel (with the dimension of Φ 2×2.2 mm, fixed on an alumina platform) and the sample of 310s steel (with the dimension of Φ 6×0.2 mm, fixed between a pair of 0.5mm thickness saphire glasses) were melted and solidified in the ultra-high temperature infrared heating furnace with the high-purity argon atmosphere. The whole experiment equipment is shown in Figure S4. During HE-XRD test, the sample of M50 bearing steel was heated and melted in the rate of 5℃/s to 1450 ℃ held for 3 minutes, solidified in the cooling rate of 1℃/s to 900℃, and was subsequently naturally cooled to the room temperature. The diffraction signals were recorded from 900 ℃ in the heating process to 900 ℃ in the cooling process. The sample of 310s steel was heated and melted in the rate of 5℃/s to 1520 ℃ held for 1 minute, solidified in the cooling rate of 0.1℃/s to 1260℃, and was subsequently naturally cooled to the room temperature. The diffraction signals were recorded from 915 ℃ in the heating process to 1260℃ in the cooling process. The sample of 321 steel was heated and melted in the rate of 5℃/s to 1450 ℃ held for 1 minute, solidified in the cooling rate of 2℃/s to 900℃, and was subsequently naturally cooled to the room temperature. The diffraction signals were recorded from 900 ℃ in the heating process to 900 ℃ in the cooling process. The data of HE-XRD test was processed by the Fit-2D software.

**Figure captions**

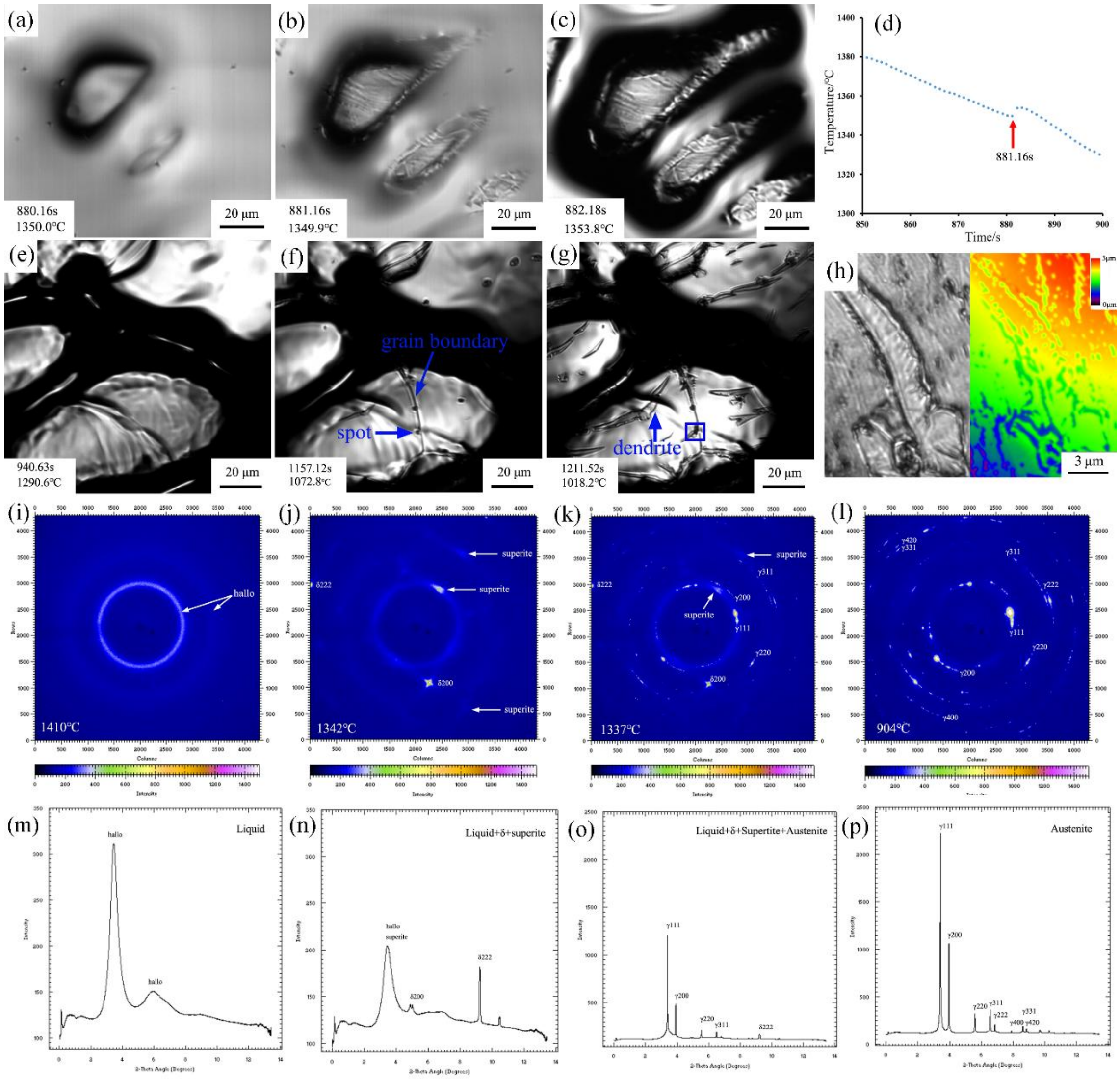

Figure 1 The phase transformation in-situ observed on the surface and HE-XRD test of the M50 steel. (a)-(c) the abrupt superite to austenite transition within initial dendritic arms; (d) the temperature jump induced by the abrupt phase transition; (e)-(g) the formation of grain boundaries, segregated spots and tiny dendritic segregation zone on the surface of initial dendritic arms; (h) the 3D morphology of the tiny dendritic segregation zone; (i) and (m) the XRD pattern and peaks of liquid at 1410℃; (j) and (n)

the XRD pattern and peaks before the superite to austenite transition at 1342℃; (k) and (o) the XRD pattern and peaks during the superite to austenite transition at 1337℃; (p) and (t) the XRD pattern and peaks after the superite to austenite transition tested at 904℃.

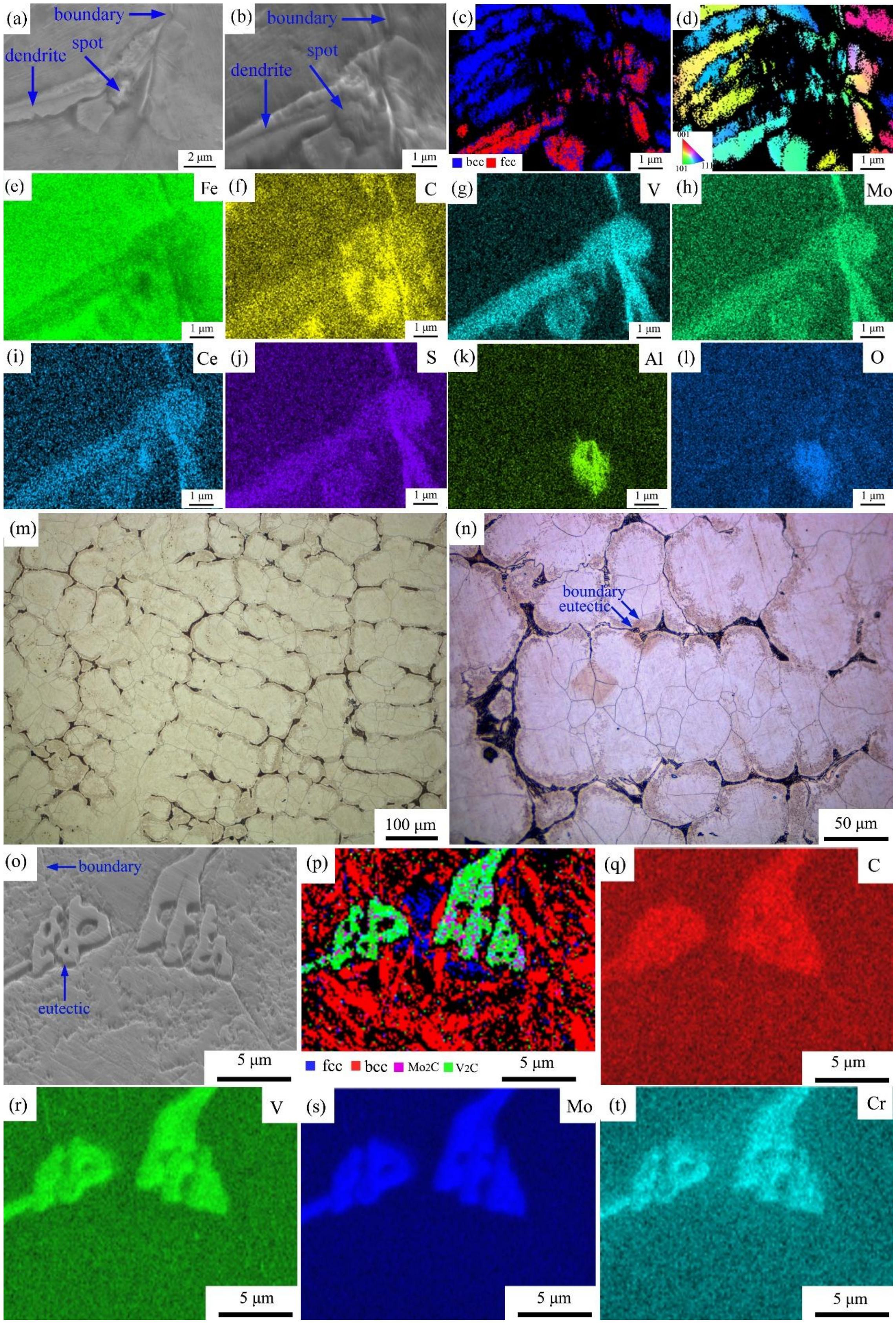
(a)
boundary
dendrite
spot
2 μm
(b)
boundary
dendrite
spot
1 μm
(c)
bcc
fcc
1 μm
(d)
1 μm
(e)
Fe
1 μm
(f)
C
1 μm
(g)
V
1 μm
(h)
Mo
1 μm
(i)
Ce
1 μm
(j)
S
1 μm
(k)
Al
1 μm
(l)
O
1 μm
(m)
100 μm
(n)
boundary
eutectic
50 μm
(o)
boundary
eutectic
5 μm
(p)
fcc
bcc
Mo2C
V2C
5 μm
(q)
C
5 μm
(r)
V
5 μm
(s)
Mo
5 μm
(t)
Cr
5 μm

Figure 2 The solidification microstructure of the M50 steel. (a) the morphologies of grain boundaries, spot segregations and tiny dendritic segregation zone amplified in the square of Figure 1(g); (b)-(d) the phase distribution and orientation of the dendritic segregation zone and the rest base; (e)-(l) the solute segregation in the dendritic segregation zone and segregated spot; (m) the dendritic pattern in the interior solidification microstructures; (n) the interdendritic eutectic and the dendritic arm composed of original austenite grains; (o) the amplified eutectic and original austenite grain boundaries; (p) the phase distribution in the microstructures of (o); (q)-(t) the solute segregation in the eutectic and grain boundaries.

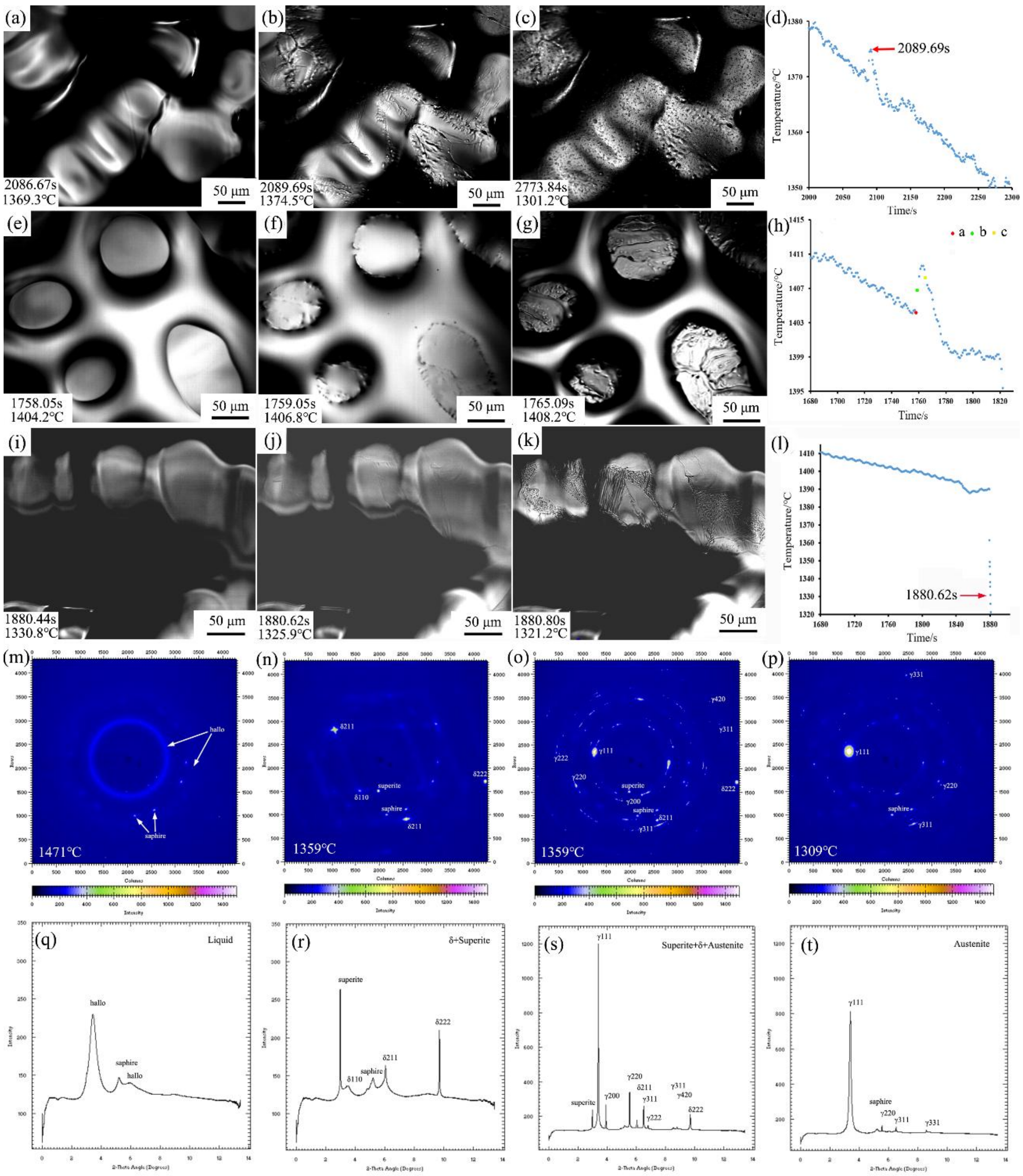

Figure 3 The phase transformation in-situ observed on the surface and HE-XRD test of the 310s stainless steel. (a)-(c) the phase transformation from superite to solid within initial dendritic arms in the 1# sample; (d) the temperature jump induced by the phase transformation in the 1# sample; (e)-(j) the phase transformation from superite to solid within initial dendritic arms in the 2# sample; (h) the cooling curve of the 2# sample; (i)-

(k) the phase transformation from superite to solid within initial dendritic arms in the 3# sample; (l) the cooling curve of the 3# sample; (m) and (q) the XRD pattern and peaks of liquid at 1471℃; (n) and (r) the XRD pattern and peaks before the superite to austenite transition at 1359℃; (o) and (s) the XRD pattern and peaks during the superite to austenite transition at 1359℃; (p) and (t) the XRD pattern and peaks after the superite to austenite transition at 1309℃.

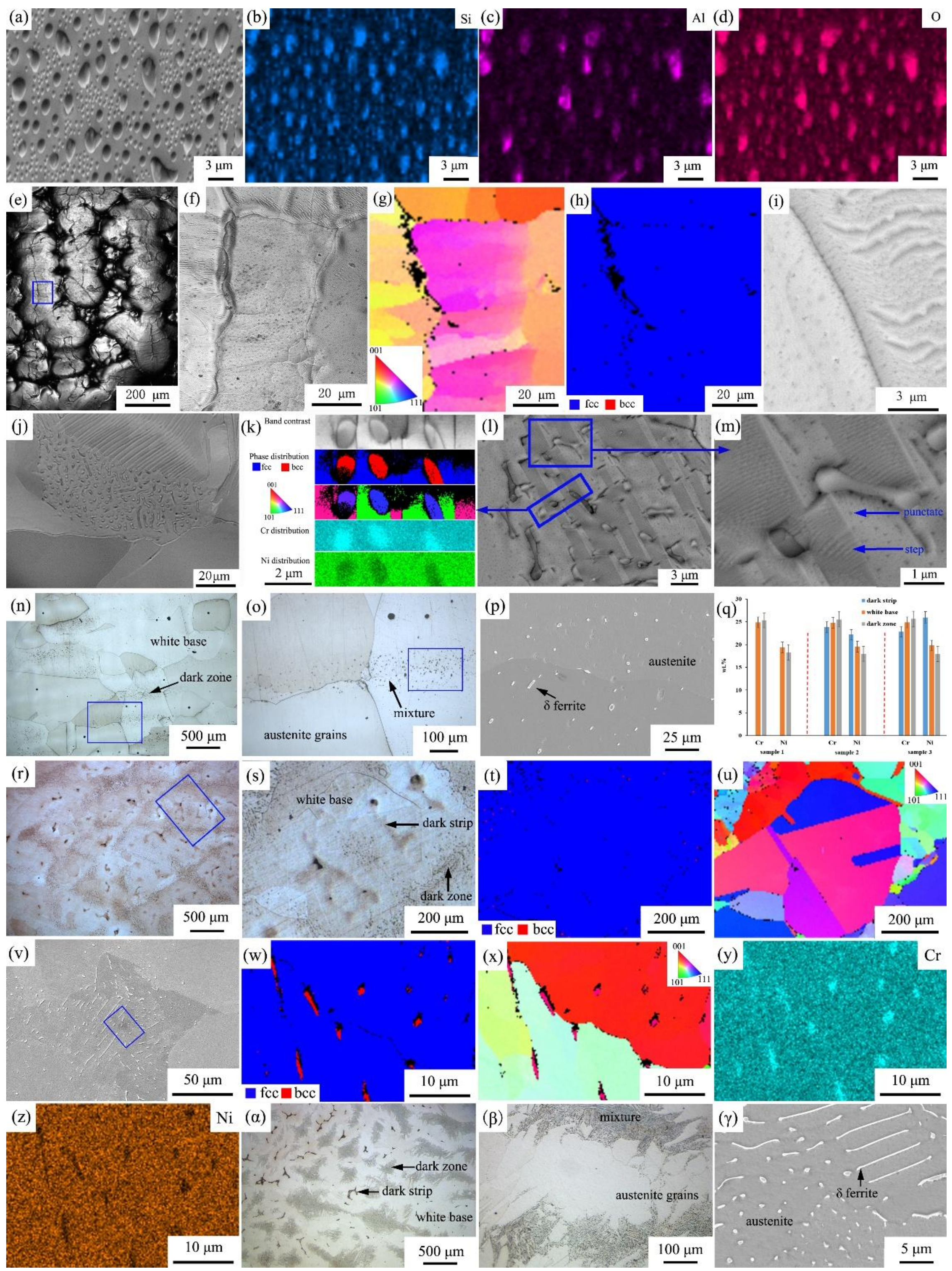

Figure 4 The solidification microstructures of the 310s steel. (a) the (Si, Al)$_x$O$_y$ compound on the surface of austenite in the 1# sample; (b)-(d) the distribution of Si, Al

and O elements in (a); (e) the solidification microstructures observed on the surface of the 2# sample; (f) the austenite grains amplified in the square within an initial dendritic arm in (e); (g) and (h) the orientation and phase distribution of austenite grains in (f); (i) the step and punctate morphologies amplified on the surface of austenite grains; (j) the austenite grains and the mixture of rod-like δ ferrite and austenite base observed on the surface of the 3# sample; (k) amplified for a square in (l), the phase distribution, orientation and segregation in the mixture of rod-like δ ferrite and austenite base; (l) the mixture of rod-like δ ferrite and austenite base; (m) amplified for a square in (l), the step and punctate morphologies on the surface of austenite with different orientations; (n) the interior solidification microstructures including coarse austenite grains and few dark zones in the 1# sample; (o) coarse austenite grains and few dark zones amplified in the 3# sample; (p) granular δ ferrite and austenite base amplified in the dark zone of 3# sample; (p) the segregation between the dark strip, white base and dark zone in three samples; (r) the interior solidification microstructures in the 2# sample; (s) the dark strips, the white base and dark zones amplified for the square in (r); (t) and (u) the phase distribution and orientation of microstructures in (s); (v) the rod-like δ ferrite and austenite base amplified in the dark zone of 1# sample; (w) and (x) the distribution and orientation of δ ferrite and austenite in the square of (v); (y) and (z) the segregation between δ ferrite and austenite in the square of (v); (α) the solidification microstructures including dark strips, the white base and dark zones in the 3# sample; (β) the separate austenite grains and interlaced boundary between the white base and the dark zone; (γ) the mixture of rod-like δ ferrite and austenite base amplified in the dark zone.

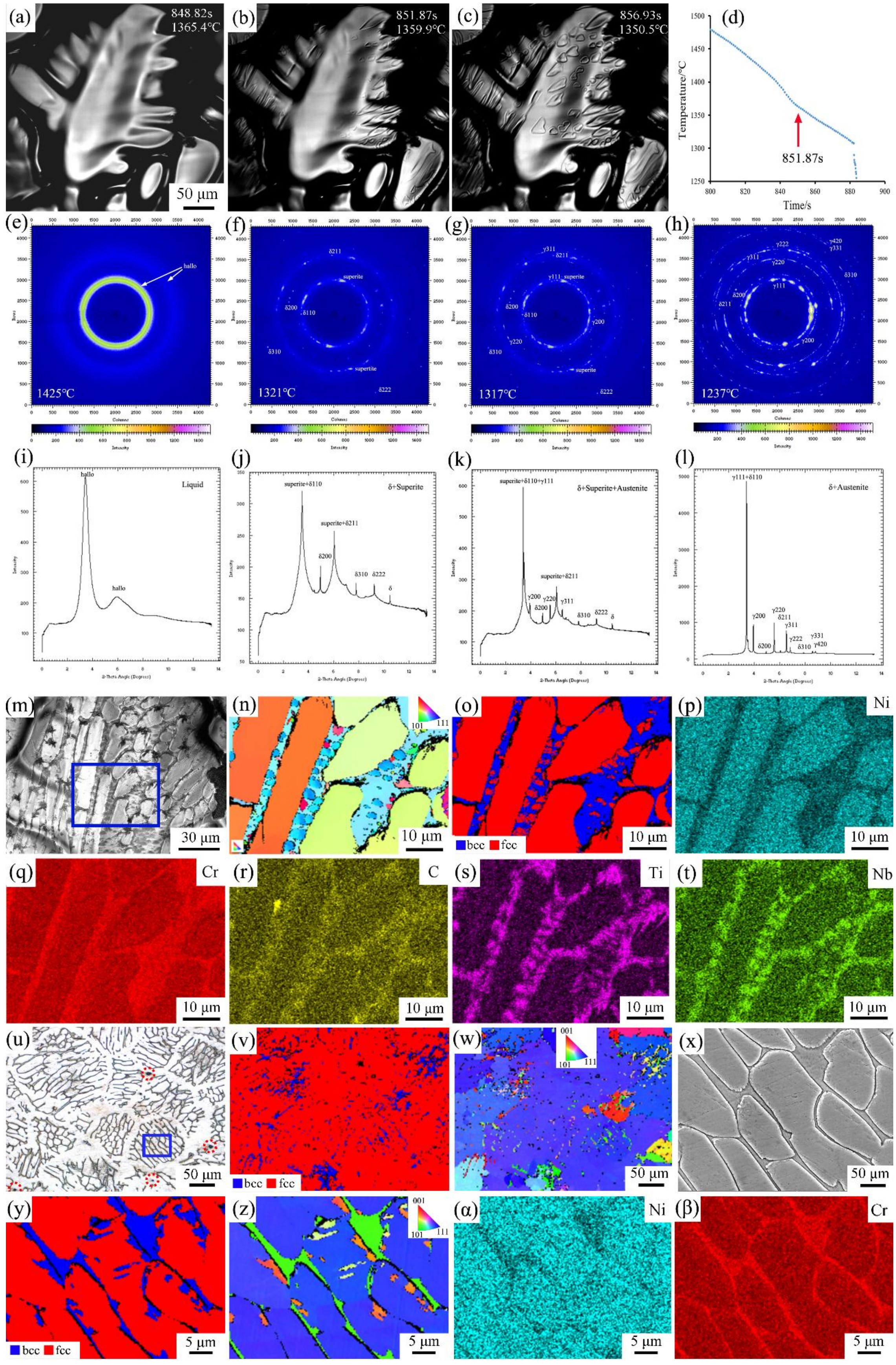

(a)
848.82s
1365.4°C
50 μm
(b)
851.87s
1359.9°C
(c)
856.93s
1350.5°C
(d)
Temperature/°C
Time/s
851.87s
(e)
hallo
1425°C
(f)
1321°C
(g)
1317°C
(h)
1237°C
(i)
Liquid
(j)
δ+Superite
(k)
δ+Superite+Austenite
(l)
δ+Austenite
(m)
30 μm
(n)
10 μm
(o)
bcc
fcc
(p)
Ni
(q)
Cr
(r)
C
(s)
Ti
(t)
Nb
(u)
(v)
(w)
(x)
(y)
5 μm
(z)
(α)
(β)

Figure 5 The phase transformation and solidification microstructures of the 321 stainless steel. (a)-(c) the in-situ morphology evolution from the superite to austenite blocks within initial dendrites; (d) the cooling curve during the phase transformation; (e) and (i) the XRD pattern and peaks of liquid at 1425℃; (f) and (j) the XRD pattern and peaks before the superite to austenite transition at 1321℃; (g) and (k) the XRD pattern and peaks during the superite to austenite transition at 1317℃; (h) and (l) the XRD pattern and peaks after the superite to austenite transition at 1237℃; (m) the solidification microstructures on the surface; (n) and (o) the orientation and distribution of austenite and δ ferrite in the square of (m); (p)-(t) the solute segregation within microstructures in the square of (m); (u) the austenite network containing porosities (labelled in the red circles) and the mixture of dendritic austenite and interstitial δ ferrite observed in the sectioned sample; (v) and (w) the distribution and orientation of austenite and δ ferrite in (u); (x) the dendritic austenite and interstitial δ ferrite amplified in the square of (u); (y) and (z) the distribution and orientation of austenite and δ ferrite in (x); (α) and (β) the solute segregation between austenite and δ ferrite in (x).

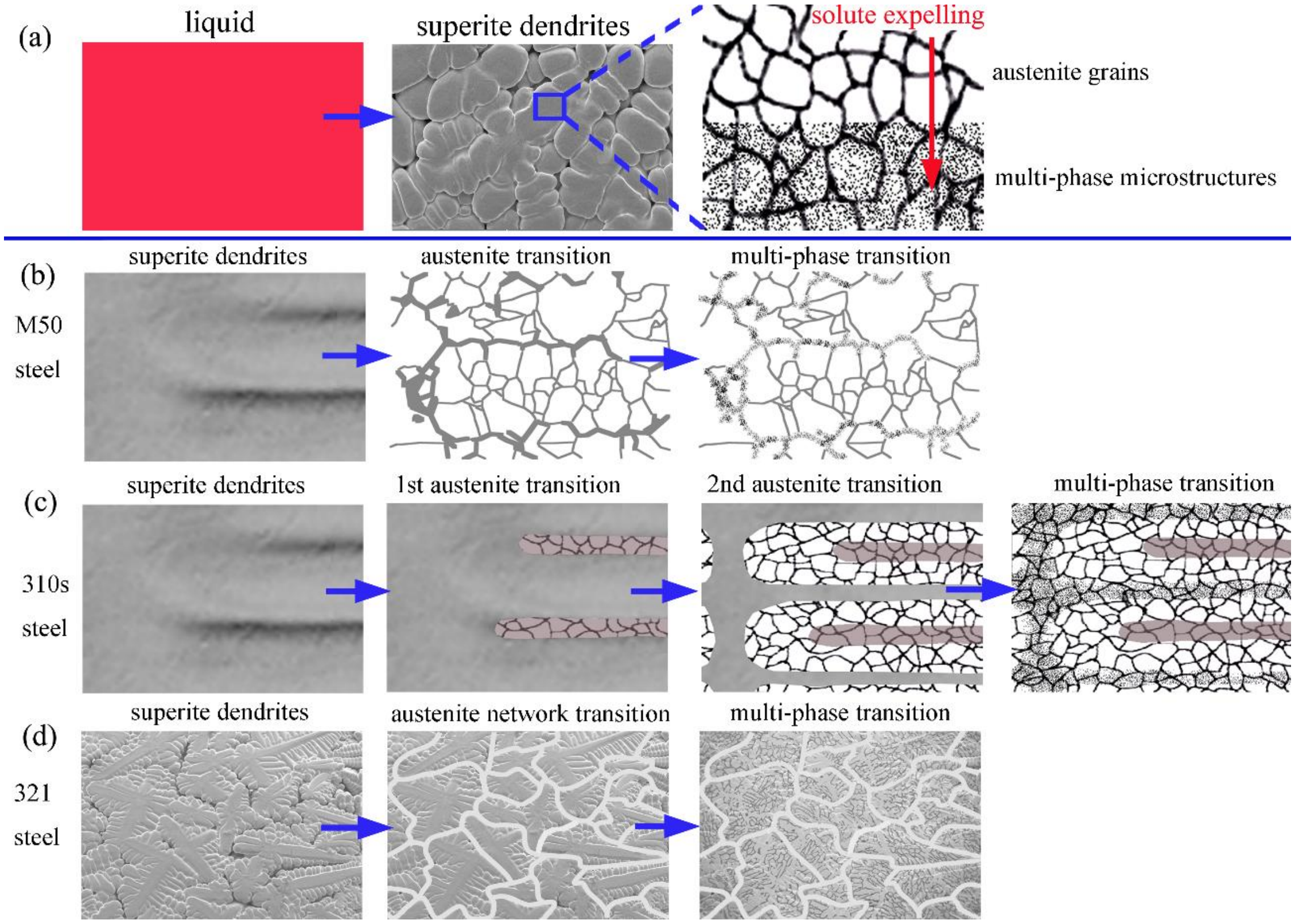

Figure 6 The sketch for phase transformation process during steel solidification. (a) a general transition process from liquid to superite dendrites and the final austenite grains and multi-phase microstructures; (b) the sketch for the phase transformation from superite phase to solid phase in M50 steel, in which most superite phase transits to austenite grains, and a little residual superite phase transits to eutectic; (c) the sketch for the phase transformation from superite phase to solid phase in the 310s steel, in which the transition from superite phase to austenite grains happens in two distinct steps, and the residual superite phase transits to mixed austenite and rod-like δ ferrite; (d) the sketch for the phase transformation from superite phase to solid phase in the 321 steel, in which a little superite phase on the boundary of dendritic grains transits to a network of austenite, and most superite phase transits to austenite grains and interstitial δ ferrite.

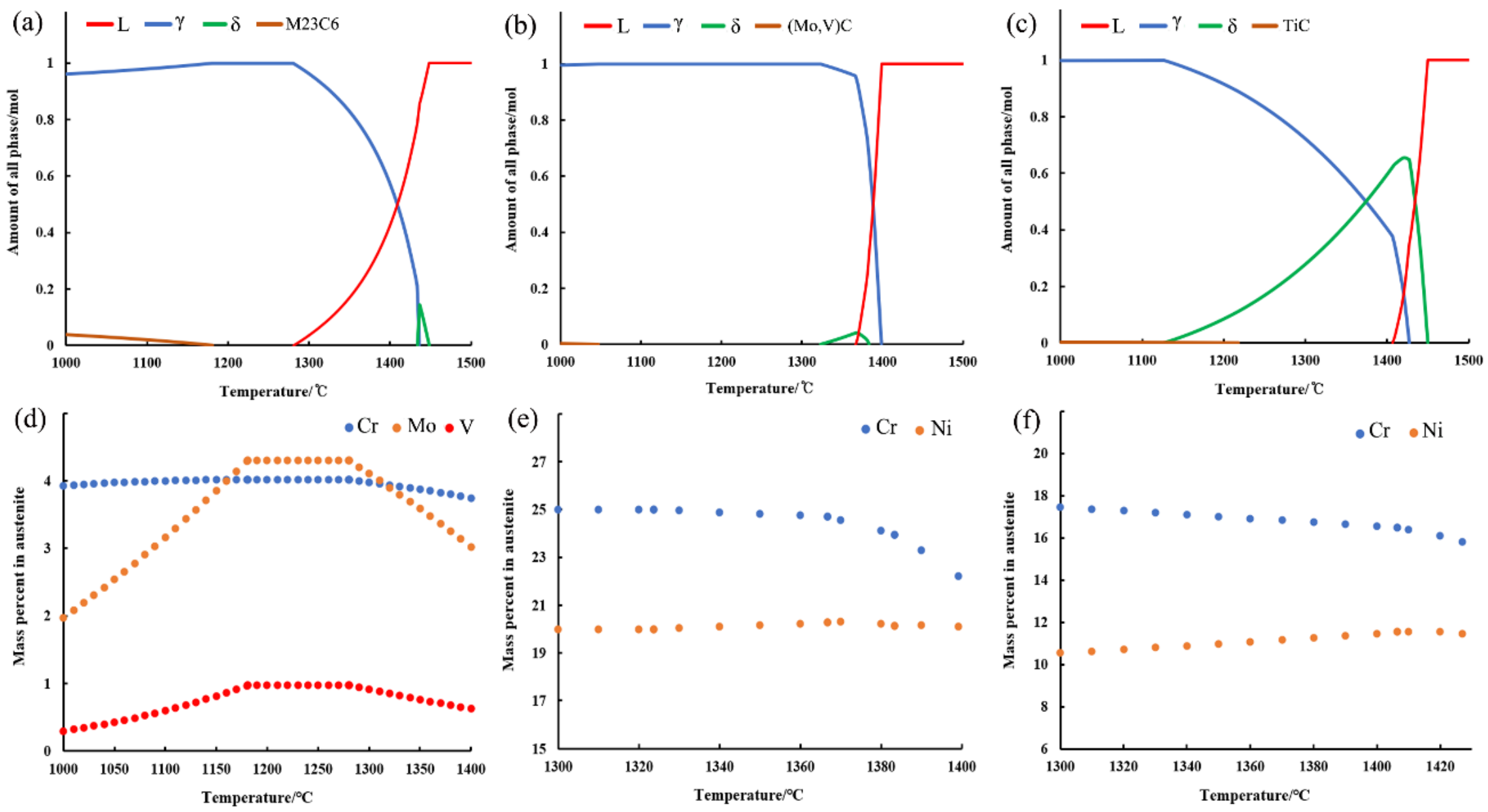

Figure S1 the phase transformation process and solute content in austenite during equilibrium solidification calculated by Thermos-Calc software. (a) the phase transformation process of the M50 steel; (b) the phase transformation process of the 310s steel; (c) the phase transformation process of the 321 stainless steel; (d) the solute content of austenite versus temperature variation in the M50 steel; (e) the solute content of austenite versus temperature variation in the 310s steel; (f) the solute content of austenite versus temperature variation in the 321 stainless steel.

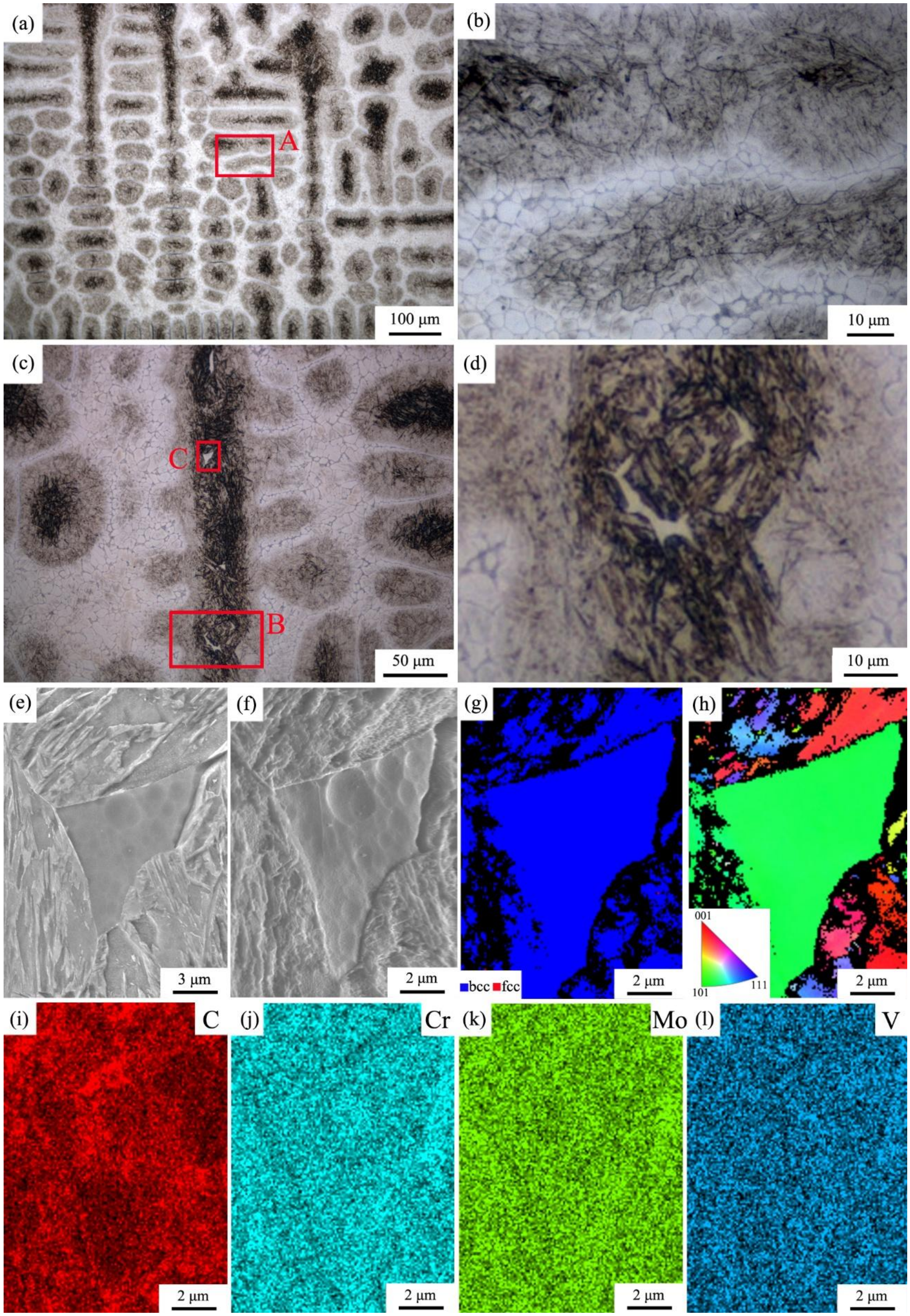
(a)
A
100 μm
(b)
10 μm
(c)
C
B
50 μm
(d)
10 μm
(e)
3 μm
(f)
2 μm
(g)
bcc
fcc
2 μm
(h)
001
101
111
2 μm
(i)
C
2 μm
(j)
Cr
2 μm
(k)
Mo
2 μm
(l)
V
2 μm

Figure S2 The microstructures of M50 steel quenched from 1370°C by the liquid GaInSn alloy. (a) the dendritic pattern reserved by quenching; (b) amplified for the A position of (a), a traditional dendritic arm is composed by many austenite grains, which finally transited to martensite; (c) some δ ferrite within an initial dendritic arm directly transited from residual superite phase by quenching; (d) the δ ferrite and surrounding martensite amplified in the B position of (c), which show previous growing austenite grains and residual superite phase; (e) the δ ferrite and surrounding martensite amplified in the C position of (c), which transited from the residual superite phase and austenite grains; (f)-(h) the phase and orientation distribution in the C position of (c) tested by EBSD mapping; (i)-(l) the distribution of C, Cr, Mo and V elements in the C position of (c) tested by EDS mapping.

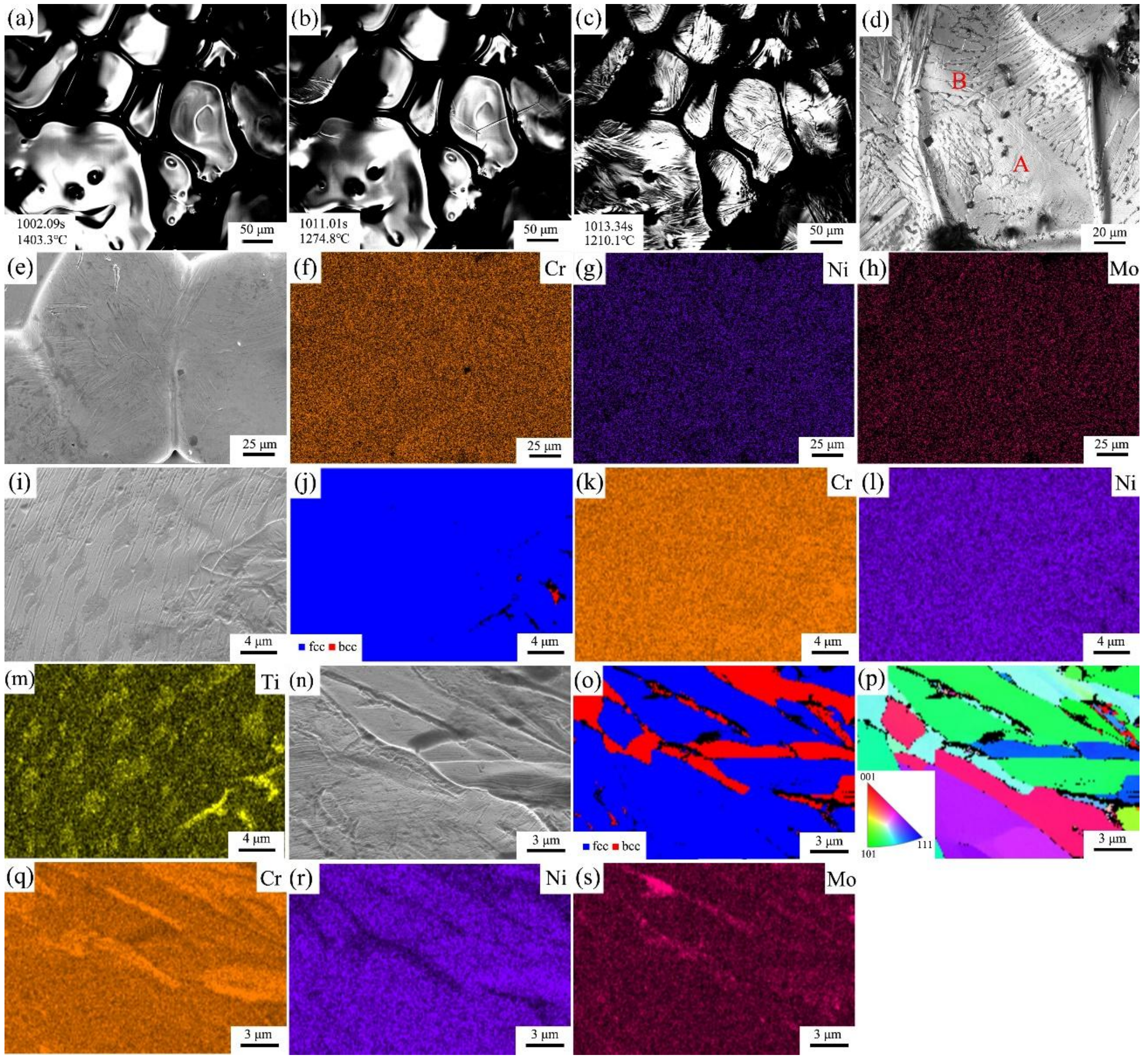

Figure S3 The phase transformation and solidification microstructures in the supplementary sample of the 321 stainless steel. (a)-(c) the in-situ observation on the transition from the superite phase to the solid during quenching process; (d) the final solidification microstructures within initial dendritic arms; (e)–(h) the negligible segregation between the center and boundary of initial dendritic arms; (i) bulk austenite transited from semi-solid in the A position of (d); (j) the phase mapping of (i) by EBSD; (k) (l) and (m) the distribution of Cr, Ni and Ti elements in (i) by EDS mapping; (n) austenite strips and interstitial δ ferrites with different orientations transited from superite

in the B position of (d); (o) and (p) the phase and orientation mapping of (n) by EBSD; (q)-(s) the obvious segregation of Cr, Ni and Mo elements in (n) by EDS mapping.

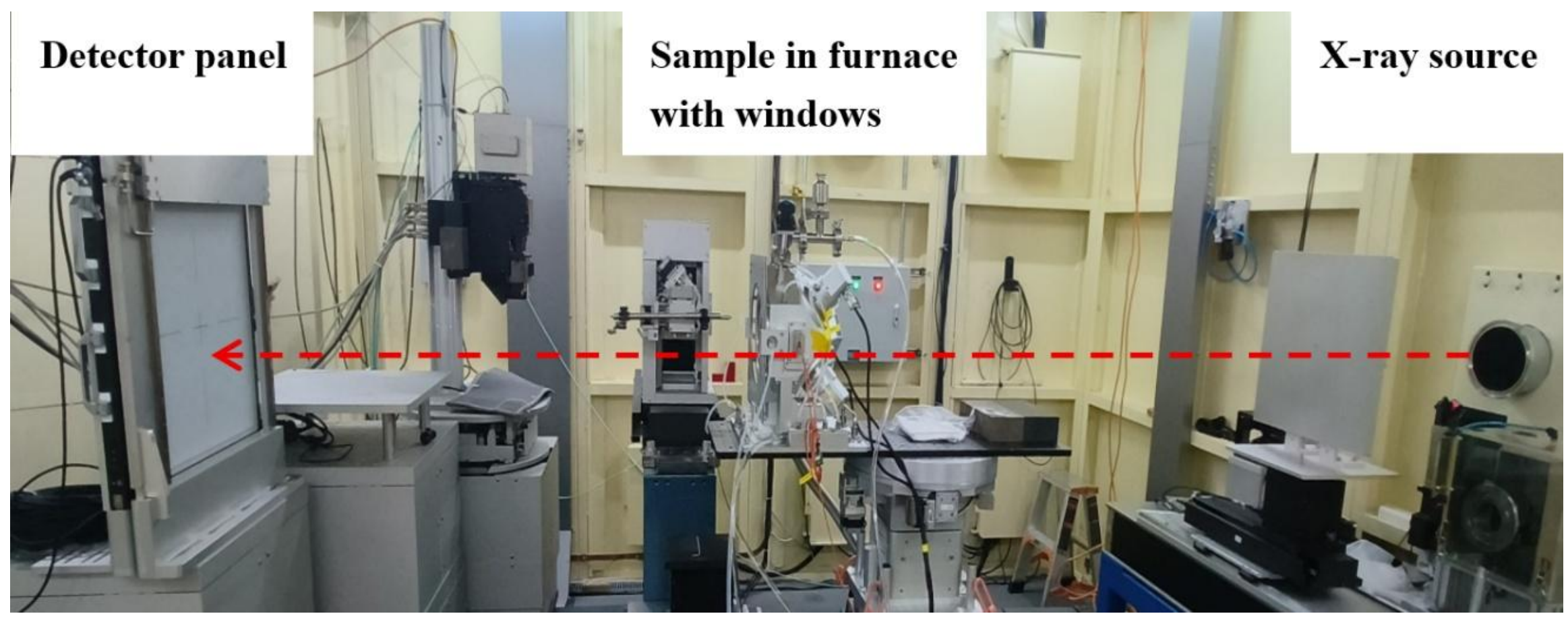

Figure S4 The experiment equipment in HE-XRD test.